\documentclass[prl,floatfix,showpacs,amsmath,twocolumn]{revtex4}
\usepackage{amssymb}
\usepackage{graphicx}

\begin{document}

\newcommand{\be}{\begin{eqnarray}}
\newcommand{\ee}{\end{eqnarray}}
\newcommand{\bea}{\begin{eqnarray}}
\newcommand{\eea}{\end{eqnarray}}
\newcommand{\bma}{\begin{subequations}}
\newcommand{\ema}{\end{subequations}}
\def\lR{l^2_{\mathbb{R}}}
\def\RR{\mathbb{R}}
\def\E{\mathbf e}
\def\D{\boldsymbol \delta}
\def\S{{\cal S}}
\def\T{{\cal T}}
\def\dd{\delta}
\def\one{{\bf 1}}

\title{DMRG and periodic boundary conditions: a quantum information
perspective}

\author{F. Verstraete, D. Porras, and J. I. Cirac}
\affiliation{Max-Planck Institut f\"ur Quantenoptik,
Hans-Kopfermann-Str. 1, Garching, D-85748, Germany}

\date{\today}

\begin{abstract}
We introduce a picture to analyze the density matrix
renormalization group (DMRG) numerical method from a quantum
information perspective. This leads us to introduce some
modifications for problems with periodic boundary conditions in
which the results are dramatically improved. The picture also
explains some features of the method in terms of entanglement and
teleportation.
\end{abstract}

\maketitle

The discovery and development of the DMRG method \cite{WhitePRL,WhitePRB} to treat
quantum many--body systems has enabled us to analyze and understand the physical
properties of certain condensed matter systems with unprecedent precision
\cite{DMRGBook}. Originally envisioned for 1D systems with short--range
interactions at zero temperatures, during the last years this method has been
successfully extended to other situations \cite{DMRGBook}. Its mathematical
foundations have been established \cite{RomerPRL,MiguelAngel} in terms of the
so--called matrix product states (MPS) \cite{Fannes} and by now there exists a
coherent theoretical picture of DMRG.

At the same time, the field of Quantum Information Theory (QIT) has emerged to
describe the properties of quantum many--body systems from a different point of
view. A theory of entanglement has been established, and has allowed us to
describe and understand phenomena like teleportation \cite{teleportation}, and to
use them in the fields of communication and computation \cite{NielsenChuang}.
Recently it has been shown that QIT may also shed some new light in our
understanding of condensed matter systems \cite{Vidal,Verstraete}, and, in
particular, in the DMRG method \cite{Osborne,Korina}.

In this work we analyze the standard DMRG method using a physical
picture which underlies QIT concepts. The picture has its roots in
the AKLT model \cite{AKLT} and allows us to understand why DMRG
offers much poorer results for problems with periodic boundary
conditions (PBC) than for those with open boundary conditions
(OBC), something which was realized at the origin of DMRG
\cite{WhitePRB}. It also gives a natural way of improving the
method for problems with PBC, in which several orders of magnitude
in accuracy can be gained. The importance of this result lies in
the fact that physically PBC are strongly preferable over OBC as
boundary effects are eliminated and finite size extrapolations can
be performed for much smaller system sizes.

Let us start by reviewing the simplest version of the DMRG method
for 1-D spin chains with OBC, which is typically represented as $B
\bullet B$ \cite{WhitePRB,Nishino}. We denote by $d$ the dimension
of the Hilbert space corresponding to each spin, and by $D$ the
number of states kept by the DMRG method. We assume that the spins
at the edges have dimension $d_0\ge D$ \cite{note1}. At some
particular step the chain is split into two blocks and one spin in
between. The left block ($L$) contains spins $1,\ldots,M-1$, and
the right one ($R$) spins $M+1,\ldots,N$. Then a set of $D\times
D$ matrices $A^s$ are determined such that the state
 \be
 \label{Amatrix}
 |\Psi\rangle = \sum_{s=1}^d \sum_{\alpha,\beta=1}^D
 A^s_{\alpha,\beta} |\alpha\rangle_L \otimes |s\rangle_M
 \otimes |\beta\rangle_R,
 \ee
minimizes the energy. The states $|\alpha\rangle_{L,R}$ are
orthonormal, and have been obtained in previous steps. They can be
constructed using the recurrence relations
 \be
 \label{recurrence}
 |\alpha\rangle_L = \sum_{\alpha'=1}^D \sum_{s=1}^d
 U^{[M-1],s}_{\alpha,\alpha'} |s\rangle_{M-1}\otimes
 |\alpha'\rangle_{L'},
 \ee
where the block $L'$ contains the spins $1,\ldots,M-2$. The new
matrices $U^{[M],s}$ are determined from $A^s$ and fulfill
 \be
 \label{condition}
 \sum_{s=1}^d U^{[M],s}\left( U^{[M],s}\right)^\dagger =\one.
 \ee
For the blocks consisting of the edge spins alone, the
$|\alpha\rangle$ are taken as the members of an orthonormal set.

In order to give a pictorial representation of the above procedure
we introduce at site $M$ two auxiliary $D$--level systems, $a_M$
and $b_M$. The corresponding Hilbert spaces $H_{a,b}$ are spanned
by two orthonormal bases $|\alpha\rangle_{a,b}$, respectively. We
take $L$ and $a_M$ (and also $R$ and $b_M$) in the (unnormalized)
maximally entangled state
 \be
 |\phi\rangle := \sum_{\alpha=1}^D
 |\alpha\rangle\otimes |\alpha\rangle,
 \ee
We can always write
$|\Psi\rangle=P_M|\phi\rangle_{L,a_M}|\phi\rangle_{R,b_M}$, where
$P_M$ maps $H_a\otimes H_b \to H_M$, with $H_M$ the space
corresponding to the $M$--th spin and [cf. (\ref{Amatrix})]
 \be
 \label{PM}
 P_M=\sum_{s=1}^d\sum_{\alpha,\beta=1}^D A_{\alpha,\beta}^s
 |s\rangle\langle \alpha,\beta|.
 \ee
In fact, we can proceed in the same way at any other site $k\ne
1,M,N$ by defining two auxiliary systems $a_k$ and $b_k$ and a map
$Q_k$ defined as in (\ref{PM}) but with the matrices $U$ instead
of the $A$. For the edge spins $1$ and $N$ we define a single
auxiliary system $b_1$ and $a_N$, respectively and define
accordingly the operators $Q_{1,N}$ which now map $H_{b,a}\to
H_{1,N}$. Thus, the state $\Psi$ is then obtained by applying the
operators $Q_1\ldots, ,P_M,\ldots Q_{N}$ to the set of maximally
entangled states $\phi$ between the auxiliary systems $b_{k}$ and
$a_{k+1}$ ($k=1,\ldots,N-1$) [see Fig.\ 1(a)].

The DMRG procedure can be now represented as follows. At location
$M$, one finds an operator $P_M$ acting on the subsystems $a_M$
and $b_M$ by determining the matrices $A^s$. From them, one
obtains the operator $Q_M$ and goes to the next step at location
$M+1$. One proceeds in the same way, moving to the right, until
one reaches the location $N$. At that point, one starts moving to
the left until one reaches the location $1$ at which point it
moves again to the right. The procedure is continued until a fixed
point for the energy is reached, something which always occurs
since the energy is a monotonically decreasing function of the
step number. This proves that DMRG with the $B\bullet B$ is a {\em
variational method which always converges}.

\begin{figure}[t]
  \centering
  \includegraphics[width=\linewidth]{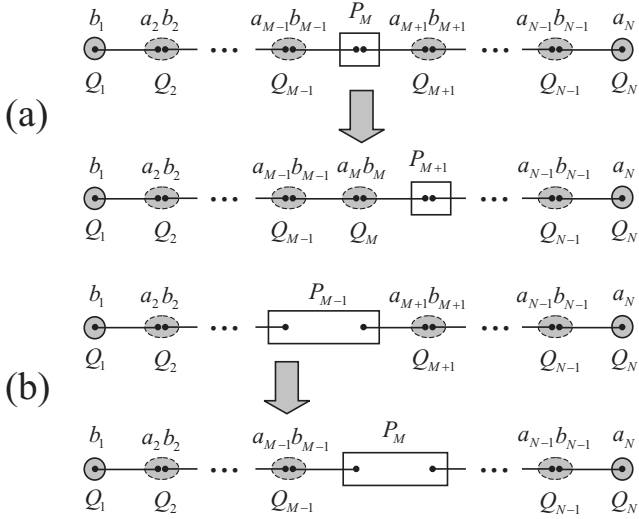}
  \caption{
  Schematic picture of the DMRG method for the $B \bullet B$ (a)
  and the $B\bullet\bullet B$ (b)
  configurations. Horizontal lines represent maximally entangled
  states $|\phi\rangle$, the ellipses and circles (squares) the
  operators $Q$ ($P$) which map the auxiliary system into the
  physical ones.}
\end{figure}

The more standard scenario ($B\bullet\bullet B$) is represented in
Fig.\ 1(b). The operator $P_M$ acts on the auxiliary subsystems
$a_{M}$ and $b_{M+1}$ and maps $H_a\otimes H_b\to H_{M}\otimes
H_{M+1}$. In this picture [for both configurations, Figs.\ 1
(a,b)] it is very clear that the two edge spins are treated on a
very different footing since they are represented by a single
auxiliary system which is not entangled to any other.

In the case of a problem with PBC a slight modification of the
scheme is used \cite{WhitePRB}. The idea is to still separate the
system into two blocks and two spins as before but now with the
configuration $B\bullet B\bullet$. This ensures the sparseness of
the matrices one has to diagonalize and thus it increases the
speed of the algorithm \cite{WhitePRB}. One can draw the diagram
corresponding to this procedure in a similar way as in Fig.\ 1.
The important point is that still there are always two sites (left
most and right most of both blocks $B$) which are treated
differently since they are represented by a single auxiliary spin
which is not entangled to any other. In our opinion, this is the
reason of the poor performance of the DMRG method for problems
with PBC.

\begin{figure}[t]
  \centering
  \includegraphics[width=\linewidth]{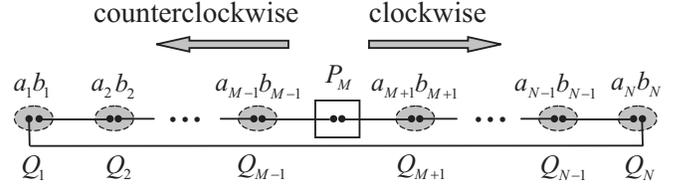}
  \label{fig2}
  \caption{
  Proposed configurations for the case of PBC. One may
  also use two spins instead of one}
\end{figure}

The method we propose is very clear in terms of this picture
(Fig.\ 2). One has to substitute {\em at all sites} $k$ the spin
by {\em two} auxiliary systems $a_k$ and $b_k$ of dimension $D$,
with $b_k$ and $a_{k+1}$ (with $a_{N+1}:=a_1$) in a maximally
entangled states and find the maps $P_k:H_a\otimes H_b\to H_k$
which lead to a state
 \be
 \label{PsiPBC}
 |\Psi\rangle= P_1\otimes P_2  \ldots P_{N-1} \otimes P_{N}
 |\phi\rangle^{\otimes N},
 \ee
with the minimal energy. This minimization can be performed in a
similar way to the one used in the standard DMRG method. Before
showing how to do this in practice, we derive some formulas in
terms of these operators. We write
 \be
 \label{PM2}
 P_k=\sum_{s=1}^d |s\rangle\langle
 \varphi_s^{[k]}|,\quad
 \langle \varphi_s^{[k]}| = \sum_{\alpha,\beta}
 B_{\alpha,\beta}^{[k],s} \langle \alpha,\beta|.
 \ee
Thus, the problem is solved once the states $\varphi$ (or equivalently, the
matrices $B$) are determined. Note that starting from these states, it is possible
to calculate expectation values of products of local observables \cite{RomerPRL},
since
 \be
 \label{Exp_Val1}
 \langle \Psi|O_1\ldots O_N|\Psi\rangle = {\rm Tr}
 \left(E^{[1]}_{O_1} \ldots E^{[N]}_{O_N}
 \right),
 \ee
where
 \be
 \label{Exp_Val2}
 E^{[k]}_{O}= \sum_{s,s'=1}^{d} \langle s|O|s'\rangle
 B^{[k],s}\otimes \left(B^{[k],s}\right)^\ast.
 \ee
Thus, the main idea to perform the minimization is very simple.
Given the Hamiltonian $H$ describing the system, one chooses one
site $M$ and writes the energy as
 \be
 E=\frac{\langle\Psi|H|\Psi\rangle}{\langle\Psi|\Psi\rangle} =
 \frac{\langle\psi^{[M]}|H_M|\psi^{[M]}\rangle}
 {\langle\psi^{[M]}|N_M|\psi^{[M]}\rangle},
 \ee
where $|\psi^{[M]}\rangle = \oplus_s |\psi^{[M]}_s\rangle$ is a
vector built by concatenating the $\psi^{[M]}_s$, and $N_M$ and
$H_M$ are $d\times D^2 $ hermitian square matrices which are built
using the vectors $\psi^{[k]}_s$ at $k\ne M$. For example,
$N_M=\oplus_s N_0$ is a block diagonal matrix with identical
blocks $N_0$ which has matrix elements
$(N_0)_{(\alpha,\alpha'),(\beta,\beta')} = (\tilde
N_0)_{(\alpha,\beta),(\alpha',\beta')}$, with
 \be
 \tilde N_0 = E^{[M+1]}_\one \ldots E^{[N]}_\one E^{[1]}_\one
 \ldots E^{[M-1]}_\one.
 \ee
Thus, at this step the operator $P_M$ is found by solving the
generalized eigenvalue problem
 \be
 \label{generaleigeq}
 H_M |\psi^{[M]}\rangle = \lambda N_M |\psi^{[M]}\rangle,
 \ee
with $\lambda$ minimum, which in turns gives the energy at this
step. Then one chooses another site and proceeds in the same way
until the energy converges. At the end we have all the $P_k$ and
can evaluate all expectation values.

The above method is not very efficient numerically. First, the
matrix $N_0$ may be ill conditioned. Second, one stores many
matrices ($\sim N^2$) and performs many matrix multiplications
($\sim N^2$) at each step. Now we explain how one can make the
method much more efficient.

Let assume that we have a set of spins in a ring. The idea is to
determine operators $P_k$ in a clockwise order (first $P_1$, then
$P_2$, until $P_{N-1}$), then improve them following a
counterclockwise ordering (from $P_{N}$ to $P_2$), then again
clockwise, until the fixed point is reached. At each step, a
normalization condition similar to (\ref{condition}) is imposed,
depending on whether we are in a clockwise or counterclockwise
cycle, which makes the matrix $N_M$ well behaved. On the other
hand, at each step only the operators which are strictly needed in
later steps are calculated in an efficient way and stored.

The normalization condition is based on the following fact. Given
the state $\Psi$, characterized by matrices $B$, if we substitute
$B^{[M],s}\to B^{[M],s} X:=U^{[M],s}$ and $B^{[M+1],s}\to
X^{-1}B^{[M+1],s}$, where $X$ is a nonsingular matrix, we obtain
the same state. Analogously, we can substitute $B^{[M],s}\to Y
B^{[M],s}:=V^{[M]}$ and $B^{[M-1],s}\to B^{[M-1],s} Y^{-1}$. We
choose $X$ in the clockwise cycles to impose (\ref{condition}) and
$Y$ in the counterclockwise ones to impose
 \be
 \sum_{s=1}^d \left(V^{[M],s}\right)^\dagger\left( V^{[M],s}\right) =\one.
 \ee
Thus, at the point of determining the operator $P_M$,
 \be
 |\Psi\rangle= Q_1\otimes \ldots Q_{M-1} \otimes P_M \otimes \tilde Q_{M+1} \ldots
 \otimes \tilde Q_{N}|\phi\rangle^{\otimes N},
 \ee
where $Q_k$ and $\tilde Q_k$ are defined as in (\ref{PM2}) but
with $U$ and $V$ instead of $B$, respectively. Thus, the operators
$X$ and $Y$ are all of them moved over, such that they are now
included in those corresponding to $P_M$. It can be easily shown
that these conditions on the operator $U$ ($V$) are equivalent to
imposing that $E_\one$ has the maximally entangled state
$|\phi\rangle$ as right (left) eigenvector with eigenvalue 1. This
is immediately reflected in the fact that the matrix $N_M$ is
better behaved, which makes the problem numerically stable.

Let us now illustrate how the procedure works with simplest
nearest neighbor Hamiltonian $\sigma_z^{[k]}\sigma_z^{[k+1]}$,
namely the Ising Model. Let us assume that we are running the
optimization of the operators clockwise and that we want to
determine $P_M$. So far, in previous steps, apart from the
matrices $U$ and $V$, we have stored: (a) For each $k<M$, the
following four operators:
 \bma
 \label{mat}
 \begin{eqnarray}
 r_k&:=&E_\one^{[1]} E_\one^{[2]} \ldots E_\one^{[k-2]}E_\one^{[k-1]},\\
 s_k&:=&E_{\sigma_z}^{[1]} E_\one^{[2]} \ldots E_\one^{[k-2]}E_\one^{[k-1]},\\
 t_k&:=&E_\one^{[1]} E_\one^{[2]} \ldots E_\one^{[k-2]} E_{\sigma_z}^{[k-1]},\\
 h_k&:=&\sum_{n=1}^{k-2} E_\one^{[1]} E_\one^{[2]}\ldots
 E_{\sigma_z}^{[n]}
 E_{\sigma_z}^{[n+1]} \ldots E_\one^{[k-2]} E_\one^{[k-1]},
 \end{eqnarray}
 \ema
(b) For each $k>M$ other four similar operators which contain
products from $E^{[k]}$ to $E^{[N]}$. With them, one can build
$H_M$ and $N_0$ by few matrix multiplications and thus determine
$P_M$ by solving (\ref{generaleigeq}). From it, $Q_M$ is
determined. Then, we construct $r_{M+1},s_{M+1},t_{M+1}$ and
$h_{M+1}$ starting from $r_M,s_M,t_M$ and $h_M$ \cite{note2}. We
continue in the same vein, finding four matrices at each step, and
storing them, until we reach $N$. Then we start moving
counterclockwise and start constructing the corresponding four
matrices at each step. Notice that in order to construct the
matrices $H_M$ and $N_0$ we will have to use the stored matrices
(\ref{mat}) which were determined when we were moving clockwise.
Thus, with this procedure we have to store of the order of $4N$
matrices of dimension $D^2$ (apart from the matrices $U$, $V$ and
the last $B$'s) but the number of operations per step is
independent of $N$. At the end, when we have reached the fixed
point, we can determine the expectation value of any operator by
using (\ref{Exp_Val1}) and determining the required matrices using
(\ref{Exp_Val2}). Note that if the problem has translational
symmetry, then all these evaluations are even simpler.

\begin{figure}[t]
  \centering
  \includegraphics[width=\linewidth]{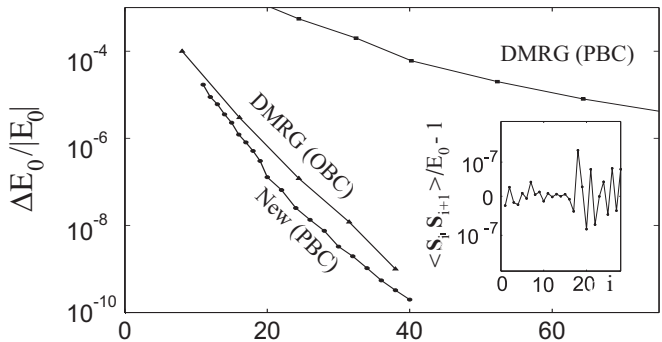}
  \label{fig1}
\caption{(Left) Comparison between DMRG (squares) \cite{WhitePRB} and the new
method
  (circles) for PBC, and $N=28$. For reference the DMRG results \cite{WhitePRB} for the Heisenberg chain with OBC
  (triangles) are also shown. (Insert) Variation of the local bond strength from the
  average along the chain, calculated with the new method and $D=40$.}
\end{figure}

We have applied the above method to the spin $1/2$ Heisenberg chain. We have
plotted in Fig.\ 3 the energies obtained as a function of $D$ and compared them
with those obtained by the standard DMRG method with OBC and PBC. From the figure
it is clear that the accuracies we obtain are comparable with those obtained with
DMRG for problems with OBC but much better than for PBC. We have determined the
errors by comparing with the exact results \cite{Exact}. In the insert of Fig.\ 3
we have plotted the local bond strength $\langle {\mathbf S}^{[k]}{\mathbf
S}^{[k+1]}\rangle$ as a function of $k$. As expected, the result is independent of
the position $k$, as opposed to what occurs with OBC.

\begin{figure}[t]
  \centering
  \includegraphics[width=\linewidth]{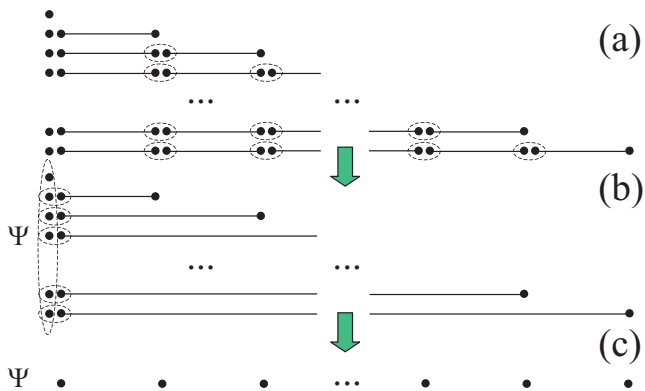}
  \label{fig4}
  \caption{
  General states can be expressed in the form (\ref{PsiPBC}).}
\end{figure}

Finally we show that the picture introduced here may be valuable to understand the
properties of states $\Psi$ in terms of the language and tools developed in the
field of QIT. First, one can easily see that the entropy of the block formed by
systems $(k_0,k_0+1,\ldots,k_1)$ is bounded by $2\log_2(D)$, as this block is
connected to the rest only via $a_{k_0}$ and $b_{k_1+1}$, and thus the rank of the
reduced density operator for the block is bounded by the product of the dimensions
of the corresponding Hilbert spaces. Secondly, it allows us to show that any state
can be written in the form (\ref{PsiPBC}) (MPS \cite{Fannes,RomerPRL}) if we
choose $D=d^N$ (actually, $D=d^{\lfloor \frac{N}{2}\rfloor}$ is sufficient). We
consider $a_k$ and $b_k$ as composed of $d$--level subsystems,
$a_k^{1},\ldots,a_k^{N}$ and $b_k^{1},\ldots,b_k^{N}$, respectively, and write
$|\phi\rangle$ as a tensor product of maximally entangled states $\phi_d$ between
$b_k^l$ and $a_{k+1}^l$. For $k=2,\ldots,N$, we choose the operators
$P_k=\one_{a_k^{k}} \otimes \langle \eta_k|$ where $|\eta_k\rangle$ is a state for
all particles but $a_k^k$, and contains $|\phi_d\rangle$ for each pair
$a_k^l$-$b_k^l$ ($l>k$) and $|0\rangle$ for the rest. The action of $P_k$ is to
teleport the entangled pairs such that at the end one has one entangled pairs
between the first system and all the rest (Fig.\ 4), while leaving all the other
auxiliary particles in $|0\rangle$. Finally, the operator $P_1$ is the product of
two operators. The first acts on particles $a_1$ and transforms $|0\rangle^{N}\to
|\Psi\rangle$. The second is $\one_{a_1^{1}} \otimes \langle \eta_1|$, where
$|\eta_1\rangle=|0\rangle_{b_1}\otimes |\phi_d\rangle^{\otimes N-1}$. This
operator first prepares the desired state $\Psi$ in particles $a_1$ and then uses
the available entangled pairs to teleport it to the the rest of the particles.

In summary, we have given a pictorial view of the DMRG method and
have identified the reason of its poor performance for problems
with PBC. Our picture immediately leads to a modified version of
the DMRG method which dramatically improves the results. This is
done at the expenses of no longer using sparse matrices, something
which limits its applications. Nevertheless, we believe that the
method may allow us to treat problems in condensed matter systems
which so far have been difficult to tackle with the standard DMRG
method. In any case, the present work illustrates how the
developments made in QIT during the last years may prove useful in
other branches of Physics.

We thank M.A. Martin-Delgado for enlightening discussions about DMRG and G.
Cabrera for sending the extended data of \cite{Exact}. Work supported by the DFG
(SFB 631), the european project and network Quprodis and Conquest, and the
Kompetenznetzwerk der Bayerischen Staatsregierung Quanteninformation.


\begin{thebibliography}{99}

\bibitem{WhitePRL}
S.R.White, Phys. Rev. Lett. {\bf 69}, 2863 (1992).

\bibitem{WhitePRB}
S.R.White, Phys. Rev. B {\bf 48}, 10345 (1992).

\bibitem{DMRGBook}
See, for example, {\it Density--Matrix Renormalization}, Eds. I.
Peschel, {\it et al.}, (Springer Verlag, Berlin, 1999).

\bibitem{RomerPRL}
S. Ostlund and S. Rommer, Phys. Rev. Lett. {\bf 75}, 3537 (1995); S. Rommer and S.
Ostlund, Phys. Rev. B {\bf 55}, 2164 (1997).

\bibitem{MiguelAngel}
J. Dukelsky {\it et al.}, Europhys. Lett. {\bf 43}, 457 (1997).

\bibitem{Fannes} M. Fannes, B. Nachtergaele and R.F. Werner, Comm.
Math. Phys. {\bf 144}, 443 (1992).

\bibitem{teleportation}
C. H. Bennett {\it et al.}, Phys. Rev. Lett. {\bf 70}, 1895
(1993).

\bibitem{NielsenChuang}
M. Nielsen and I. Chuang, {\em Quantum
Computation and Quantum Information}, Cambridge University Press
(2000).

\bibitem{Vidal}
G. Vidal {\it et al.}, Phys. Rev. Lett. {\bf 90}, 227902 (2003).

\bibitem{Verstraete}
F. Verstraete, M. Popp, and J. I. Cirac, Phys. Rev. Lett. {\bf
92}, 027901 (2004); F. Verstraete, M. A. Martin-Delgado, and J. I.
Cirac, Phys. Rev. Lett. {\bf 92}, 087201 (2004),

\bibitem{Osborne}
T. J. Osborne and M. A. Nielsen, Phys. Rev. A 66, 032110 (2002).

\bibitem{Korina}
G. Vidal, Phys. Rev. Lett. {\bf 91}, 147902 (2003).

\bibitem{AKLT}
I. Affleck {\it et al.}, Commun. Math. Phys. {\bf 115}, 477
(1988).

\bibitem{note1}
Note that this it is always possible to consider the first and
last $l$ spins as two larger spins. Here $l$ is the minimum
integer such that $d_0:=d^l\ge D$.

\bibitem{Nishino}
H. Takasaki, T. Hikihara, and T. Nishino, J. Phys. Soc. Jpn. {\bf
68}, 1537 (1999).

\bibitem{note2}
Note that this can be done more efficiently by using the fact that
the matrices $E^{[M+1]}$ have at most rank 2.

\bibitem{Exact} D. Medeiros and G.G. Cabrera, Phys. Rev. B {\bf 43}, 3703 (1991).

\end{thebibliography}
\end{document}